# Tunable Supercurrent Through Semiconductor Nanowires

Yong-Joo Doh[1,*], Jorden A. van Dam[1,*], Aarnoud L. Roest[1,2], Erik P. A. M. Bakkers[2], Leo P. Kouwenhoven[1], Silvano De Franceschi[1,**]

[1]*Kavli Institute of Nanoscience, Delft University of Technology, PO Box 5046, 2600 GA Delft, The Netherlands*
[2]*Philips Research Laboratories, Professor Holstlaan 4, 5656 AA Eindhoven, The Netherlands*
[*]*These authors contributed equally to this work*
[**]*Present address: Laboratorio Nazionale TASC-INFM, I-34012 Trieste, Italy*

**Nanoscale superconductor-semiconductor hybrid devices are assembled from InAs semiconductor nanowires individually contacted by aluminum-based superconductor electrodes. Below 1 K, the high transparency of the contacts gives rise to proximity-induced superconductivity. The nanowires form superconducting weak links operating as mesoscopic Josephson junctions with electrically tunable coupling. The supercurrent can be switched on/off by a gate voltage acting on the electron density in the nanowire. A variation in gate voltage induces universal fluctuations in the normal-state conductance which are clearly correlated to critical current fluctuations. The ac Josephson effect gives rise to Shapiro steps in the voltage-current characteristic under microwave irradiation.**

The recent interest for chemically grown semiconductor nanowires arises from their unique versatility which translates into a wide range of potential applications. Many important proofs of concept have already been provided such as field effect transistors, elementary logic circuits, resonant tunneling diodes, light emitting diodes, lasers, and biochemical sensors (*1-3*). These achievements, together with the recent advance in the monolithic integration of III-V nanowires with standard Si technology (*4,5*), yield great promise for the development of next-generation (opto-)electronics. Simultaneously, the high degree of freedom in nanowire growth and device engineering creates new opportunities for the fabrication of controlled one-dimensional systems for low-temperature applications and fundamental science. Quantum confinement and single-electron control have been achieved in a variety of single-nanowire devices (*6-8*). In these experiments the transport properties were dominated by Coulomb interactions among conduction electrons due to the presence of high-resistance barriers either at the interface with the metal leads or within the nanowire itself.

Here we address an entirely different regime in which the nanowires are contacted by superconducting electrodes with deliberately low contact resistance. While Coulomb blockade effects are suppressed, the semiconductor nanowires acquire superconducting properties due to the proximity effect, a well-known phenomenon which can be described as the leakage of Cooper pairs of electrons from a superconductor (S) into a normal-type conductor (N) (*9*). The proximity effect takes place only if the S-N interface is highly transparent to electrons. This requirement is particularly hard to meet when the N element is a semiconductor, the major obstacle being posed in most cases by the unavoidable presence of a Schottky barrier. In this respect, indium arsenide (InAs) is an exceptional semiconductor as it can form Schottky-barrier-free contacts with metals (*10*). This motivates our choice for this material in the present work.

The InAs nanowires are grown via a catalytic process based on a vapor-liquid-solid mechanism (*11*). The nanowires are monocrystalline with diameters in the 40 – 130 nm range and lengths of 3 – 10 μm. From field-effect electrical measurements (discussed below), we find *n*-type conductivity with an average electron density $n_s = (2–10) \times 10^{18}$ cm$^{-3}$, and an electron mobility $\mu = 200 – 2000$ cm$^2$/Vs. These values correspond to a mean free path, $l = 10 – 100$ nm. Right after growth, the nanowires are transferred to a p$^+$ silicon substrate with a 250-nm-thick SiO$_2$ overlayer. The conductive substrate is used as a back gate to vary the electron density in the nanowires. Custom metal electrodes are defined by e-beam lithography followed by e-beam evaporation of Ti(10 nm)/Al(120 nm). Prior to metal deposition the nanowire surface is deoxidized by a 6 s wet etching in buffered hydrofluoric acid. No thermal annealing is performed in order to minimize inter-diffusion at the contacts. The spacing, *L*, between the source and drain electrodes is varied between 100 and 450 nm. To perform four-point measurements, both source and drain electrodes are split in two branches (Fig. 1A). A representative single-nanowire device is shown in Fig. 1B.

The nanowire devices exhibit Ohmic behavior with a normal state resistance, $R_N$, in the 0.4 – 4 kΩ range. $R_N$ is virtually insensitive to temperature all the way down to the superconducting transition temperature of the Al-based electrodes, $T_C = 1.1$ K. Below $T_C$, the proximity effect manifests itself through the appearance of a dissipation-less supercurrent which can be viewed as a consequence of the diffusion of Cooper pairs throughout the entire length of the nanowire section between the two superconducting electrodes. (This requires the phase-coherence length for electron propagation in the nanowire to be longer or at least comparable to the distance between source and drain contacts.) To investigate this superconductivity regime we performed four-terminal dc measurements, in which the voltage across the nanowire, $V$, is measured while sweeping the bias current, $I$. In Fig. 1C we show a representative measurement taken at base temperature, T = 40 mK. The $V(I)$ characteristic exhibits a clear supercurrent branch (i.e. a region of zero-resistance) as well as a dissipative quasiparticle branch with a dependence on the sweeping direction of the source-drain current. The switching from superconductive to dissipative conduction occurs when $I$ approaches a critical current, $I_C$, leading to the abrupt appearance of a finite voltage. The reversed switching, from resistive to superconductive regime, occurs at a lower current level $I_R$. The observed $V(I)$ characteristics, as well as their hysteretic behavior, are typically found in capacitively shunted Josephson junctions (*12*). (In our devices, the shunting capacitors are formed between the source/drain electrodes and the conductive Si substrate.)

The presence of a supercurrent has been assessed in 14 devices (90% yield) with critical currents ranging from a few nA to 135 nA at 40 mK (see Fig. 2S for the temperature dependence). On average, $I_C$ decreases with $R_N$ (see inset to Fig. 1C), and the $I_C R_N$ product, a typical figure of merit for Josephson junctions, varies between 2 – 60 µV. The highest value is comparable to the expectation for an ideal S-N-S junction embedding a short and diffusive ($l<L$) normal conductor, i.e. $I_C R_N \sim \Delta_0/e$ where $\Delta_0$ = 100–120 µeV is the superconducting energy gap of the contact electrodes obtained from finite-voltage measurements discussed below. The semiconductor nature of the nanowires allows the magnitude of the critical current to be controlled by a voltage, $V_g$, applied to the back-gate electrode (*13-17*). Due to their n-type character, a negative $V_g$ results in a reduction of the electron concentration in the nanowires leading to a higher $R_N$ and hence a lower $I_C$. Thus the nanowires can operate as tunable superconducting weak links. For sufficiently negative gate voltages, $I_C$ is entirely suppressed (Fig. 1D), where the "off" state ($I_C$ = 0) is reached for $V_g \approx -70$ V. This large voltage is due to the weak capacitive coupling between the nanowire conducting channel and the back gate. The use of alternative gating geometries, such as local top gates or gate-around configurations, would provide a much stronger coupling and, at the same time, the possibility to individually control different nanowires on the same chip. This allows a bottom-up assembly of superconducting integrated circuits based on independently addressable S-N-S elements.

To gain more insight in the Josephson behavior of the nanowire S-N-S junction, we investigate the effect of an external microwave field. Due to phase locking between the microwave angular frequency, $\omega_{rf}$, and the voltage-dependent Josephson frequency (18), $\omega_J = 2eV/\hbar$, the V(I) characteristic exhibits voltage plateaus at $V_n = n\hbar\omega_{rf}/2e$, the Shapiro steps (12) (here $\hbar$ is Plank's constant, $e$ is the electron charge, and $n=0,\pm1,\pm2,...$). A representative set of plateaux (up to $|n|=4$) is shown in Fig. 2A in the case of 4 GHz radiation. A V(I) curve for the same device in the absence of the microwave field is also shown for direct comparison. We vary the microwave frequencies in the 2 – 10 GHz range in order to verify the proportionality relation between the step height, $\Delta V$, and $\omega_{rf}$. As shown in the inset of Fig. 2A, the experimental values fall on top of the expected linear dependence $\Delta V = (\hbar/2e)\omega_{rf}$. By increasing the microwave field, proportional to the square root of the externally applied microwave power, $P_{rf}$, higher order steps become progressively visible and their step width, $\Delta I_n$, exhibits quasi-periodic oscillations. This behavior emerges clearly in a plot of differential resistance, $dV/dI$, as a function of $P_{rf}^{1/2}$ and source-drain current. In Fig. 2B we show an example of such plots corresponding to the same microwave frequency (4 GHz). The Shapiro steps appear as dark regions ($dV/dI = 0$) separated by bright (high $dV/dI$) lines corresponding to the sharp boundary between consecutive steps. The wiggling behavior of these lines reflects the quasi-periodic oscillations in $\Delta I_n$. These oscillations are quantitatively shown in Fig. 2C for $n=0$ to 4. We find good agreement with the theoretical expectation $\Delta I_n = 2I_C|J_n(2ev_{rf}/\hbar\omega_{rf})|$ (12), where $J_n$ is the $n$-th order Bessel function and $v_{rf}$ is the amplitude of rf voltage across the nanowire junction. $I_C$ and a scaling factor, $\alpha$, for the horizontal axis ($v_{rf} = \alpha P_{rf}^{1/2}$) are the only fitting parameters for all of the five theoretical curves. The observed Shapiro steps represent clear evidence of genuine Josephson coupling through the nanowire.

We now discuss in more detail the basic properties of the S-N-S nanowire devices and elucidate their mesoscopic nature. At the onset of the quasi-particle branch (i.e. just above $I_C$) the device conductance is found to be higher than the normal-state value $G_N = 1/R_N$. This is apparent from the slope of the wide-range I(V) characteristic shown in the inset of Fig. 3A (red trace). Only for $V > 0.3$ mV the slope changes to the normal-state value as becomes clear by a comparison with an I(V) trace taken at 100 mT (black trace), a magnetic field high enough to suppress superconductivity in the electrodes. Evidently, the high-bias linear I(V) at zero field does not extrapolate to the origin, as in the normal state, but to a finite excess current $I_{exc}$=0.23 µA. The enhanced conductance at low bias and the consequent excess current provide a clear indication of strong Andreev reflection (*19*) – an electron above the Fermi energy coming from the N region is reflected at the S-N interface as a phase-conjugated hole (i.e., the lack of an electron below the Fermi energy) while creating at the same time a Cooper pair in the S region. Andreev reflection at the S-N interface and phase-coherent electron propagation in the normal conductor can be viewed as the microscopic origin of the proximity effect (*10,20*). Being a two-particle process, Andreev reflection requires high interface transparency. Under this condition, it enables electrical conduction at sub-gap voltages, i.e. for $eV<2\Delta_0$, where the factor 2 accounts for the presence of two S-N interfaces in series. A sub-gap conductance, $G_{AR}$, larger than the normal-state value (up to 2 $G_N$) can be observed in the case of almost ideal interfaces having a transmission coefficient, $T_{int}$, close to unity. Shown in Fig. 3A is a plot of the differential conductance, *dI/dV*, versus *V* for the same nanowire device. We find $G_{AR}/G_N \approx 1.4$ and from the value of $I_{exc}$ we estimate $T_{int} \approx 0.75$ (*21*). Similar values are consistently obtained for most of the devices (see Fig. S4). This high transparency is in line with the best results ever achieved for micrometer-scale superconductor-semiconductor interfaces (*22*). Besides an overall conductance enhancement, the sub-gap regime is characterized by a series of peaks in *dI/dV* occurring at $V = 2\Delta_0/me$ (and *m* = 1, 2, 3) and denoted by vertical arrows. These peaks are due to multiple Andreev processes (*23-25*). Between two consecutive Andreev reflections the motion of electron-like (or hole-like) quasiparticles between the two S-N interfaces is diffusive (*l<L*) but phase coherent.

This mesoscopic character of the nanowire S-N-S junctions emerges clearly from a detailed analysis of the $V_g$-dependence. $I_C$ exhibits reproducible and time-independent fluctuations. In a color plot of *dV/dI* versus ($V_g$,*I*), shown in Fig.3B, these appear as irregular variations in the *I*-width of the zero-*dV/dI*, supercurrent branch (black region). The superimposed white trace illustrates the evolution of the corresponding normal-state conductance $G_N$ over the same $V_g$ range. Interestingly, $I_C(V_g)$ and $G_N(V_g)$ fluctuations exhibit a clear correlation which indicates a common physical origin. We interpret these $G_N(V_g)$ oscillations as universal conductance fluctuations (UCF) (*26*) associated with the phase-coherent diffusive motion along the nanowire. Similar $G_N(V_g)$ fluctuations are also found in the other nanowire devices. The corresponding rms amplitudes, $\delta G_N$, are reported in Fig. 3C. We find an average value of 0.55 $e^2/h$, which is very close to the expectation for UCF in a phase-coherent quasi one-dimensional conductor ($\delta G_N = 0.7$ $e^2/h$). $I_C$ fluctuations arising from the diffusive motion of electrons in a disordered weak link have been theoretically addressed for two distinct conditions: $\Delta_0 \ll E_{Th}$ (27) and $\Delta_0 \gg E_{Th}$ (28), where $E_{Th} = \hbar D/L^2$ is the Thouless energy in the normal conductor and $D$ is the diffusion coefficient. In the first case, which is the most pertinent to the present study (see Fig. 3C), a universal limit of $\delta I_C \sim e\Delta_0/\hbar$ is expected for the rms amplitude of the $I_C$ fluctuations, in analogy with universal conductance fluctuations in the normal state. Experimentally we find $\delta I_C = 0.2 - 3$ nA, much smaller than the expected value (~25 nA). This discrepancy could be due to a non ideal interface transparency or to an incomplete screening from the electromagnetic environment.

Finally, we would like to comment on the reproducibility of the results presented above as well as on the prospects for nanowire-based S-N structures. About 90 percent of the devices fabricated in this work have a normal-state resistance below a few kΩ and exhibit a supercurrent branch at low temperature. This indicates reproducibly low contact resistances, an important requirement for the successful up scaling to even small superconducting circuits incorporating multiple nanowire devices. The functionalities of these circuits could be integrated with conventional silicon technology or with solid-state quantum computer architectures currently under study. For instance, electrically tunable Josephson nano-junctions may be used as building blocks for superconducting quantum interference devices (SQUIDs) in which the circulating supercurrent can be switched on and off by a control voltage. Such devices may serve as switchable coupling elements between superconducting qubits (29,30). Several important issues need still to be addressed such as the local gating of individual nanowires and the replacement of aluminum with wider-gap superconductors allowing for higher operation temperatures. We believe, however, that InAs-based semiconductor nanowires can already provide a convenient basis for the development of more complex hybrid nanostructures which may enable the investigation of exotic and so far elusive phenomena resulting from the interplay between size quantization and different types of electron-electron correlations such as superconductivity, Coulomb interactions, and Kondo-type correlations.

31. Acknowledgements: J. Eroms, R. Schouten, C. Harmans, P. Hadley, Yu. Nazarov, C. Beenakker, T. Klapwijk, B. van Wees, F. Giazotto. Financial support was obtained from the Dutch Fundamenteel Onderzoek der Materie (FOM), from the Japanese Solution Oriented Research for Science and Technology (SORST) program, and from the Korean

# Figure Captions

**Fig. 1.**

(A) Schematic of a nanowire device. In four-terminal measurements current is driven between $I_+$ and $I_-$ and the voltage drop is simultaneously measured between $V_+$ and $V_-$. A gate voltage $V_g$ is applied to the p$^+$ Si substrate to vary the electron density in the nanowire.

(B) Scanning electron micrograph of a nanowire device. The nanowire diameter is determined by the size of the gold catalytic particle which is visible at the upper end of the nanowire.

(C) Voltage-vs-current, $V(I)$, characteristic for device n.1 measured in a four-terminal configuration at T = 40 mK for both increasing (red) and decreasing (blue) current bias. *Inset:* Correlation between $I_C$ and $R_N$ (the data points correspond to different devices and $V_g = 0$).

(D) $V(I)$ characteristics of device n.2 at $T$ = 40 mK for $V_g$ = 0 (red), -10 (blue), -50 (green), -60 (purple), -71 V (black). By making $V_g$ more negative the critical current is progressively reduced all the way to zero. When the supercurrent vanishes the zero-bias resistance of the device is 70 kΩ. The characteristic parameters at $V_g$ = 0 are $I_C$ = 1.2 nA and $R_N$ = 4.5 kΩ.

**Fig. 2.**

(A) $V(I)$ characteristics for device n.3 at 40 mK, with (red) and without (black) an externally applied 4-GHz radiation (this device has $R_N$ = 860 Ω and $I_C$ = 26 nA at $T$ = 40 mK). The red trace is horizontally offset by 40 nA. The applied microwave radiation results in voltage plateaus (Shapiro steps) at integer multiples of $\Delta V$=8.3µV. *Inset*: Measured voltage spacing $\Delta V$ (symbol) as a function of microwave angular frequency $\omega_{rf}$. The solid line (theory) shows the agreement with the ac Josephson relation $\Delta V = \hbar\omega_{rf}/2e$.

(B) Differential resistance, $dV/dI$, plotted on color scale as a function of the bias current, $I$, and the square root of the microwave excitation power, $P_{rf}$. In this plot the microwave frequency is fixed at 4 GHz. The voltage plateaus at $V_n = n\hbar\omega_{rf}/2e$ appear as black regions ($dV/dI$=0) labeled by the corresponding integer index, $n$. These regions are delimited by bright lines (high $dV/dI$) corresponding to the sharp increase of the $V$ between consecutive plateaus. For lower frequencies we have also observed half-integer steps, i.e. $n = \pm1/2, \pm3/2,\ldots$ (see Fig. S3).

(C) Current width $\Delta I_n$ of the *n*-th Shapiro steps (*n*=0,1,2,3,4) versus $P_{rf}^{1/2}$ as extracted from (B).

The five solid lines are Bessel-type functions obtained from a single two-parameter fit (see text).

**Fig. 3.**

(A) V-dependence of the differential conductance, $dI/dV$, normalized to the normal-state value $R_N^{-1}$ for device n. 1. The vertical arrows indicate three $dI/dV$ maxima at $V_m = 2\Delta_0/me$ ($m$ = 1, 2, 3) due to multiple Andreev reflection. *Inset*: $I(V)$ characteristics at 40 mK for zero magnetic field (red trace) and for 100 mT field perpendicular to the substrate (black trace). The supercurrent branch at zero field is not clearly visible in this large bias range.

(B) Differential resistance, $dV/dI$, vs. bias current, $I$, and gate voltage, $V_g$, for device n.2. As $V_g$ is varied from 0 to negative values, the supercurrent branch (black region) shrinks and simultaneously exhibits reproducible fluctuations. These fluctuations correlate with those of the normal-state conductance, $G_N(V_g)$, as it is shown by the comparison with the superimposed plot (white line) sharing the same horizontal scale. (In practice, $G_N(V_g)$ is obtained by measuring $dV/dI$ at $V>2\Delta_0/e$.)

(C) The Thouless energy, $E_{Th}$, and the rms value, $\delta G_N$, of the $G_N(V_g)$ fluctuations are plotted as a function of the distance, $L$, between source and drain contacts. $E_{Th} = \hbar D/L^2$, where $D = l\, v_F/d$ is the diffusion coefficient for a $d$-dimensional conductor expressed in terms of the mean-free path, $l$, and the Fermi velocity, $v_F$. $l$ and $v_F$ are extracted from the mobility and the carrier density obtained from the pinch-off $G_N(V_g)$ characteristics (*6*). Since $l$ is typically comparable to the nanowire diameter, we take $d=3$ instead of $d=1$. Even with this conservative assumption, the Thouless energy $\geq \Delta_0$.

# Fig. 1

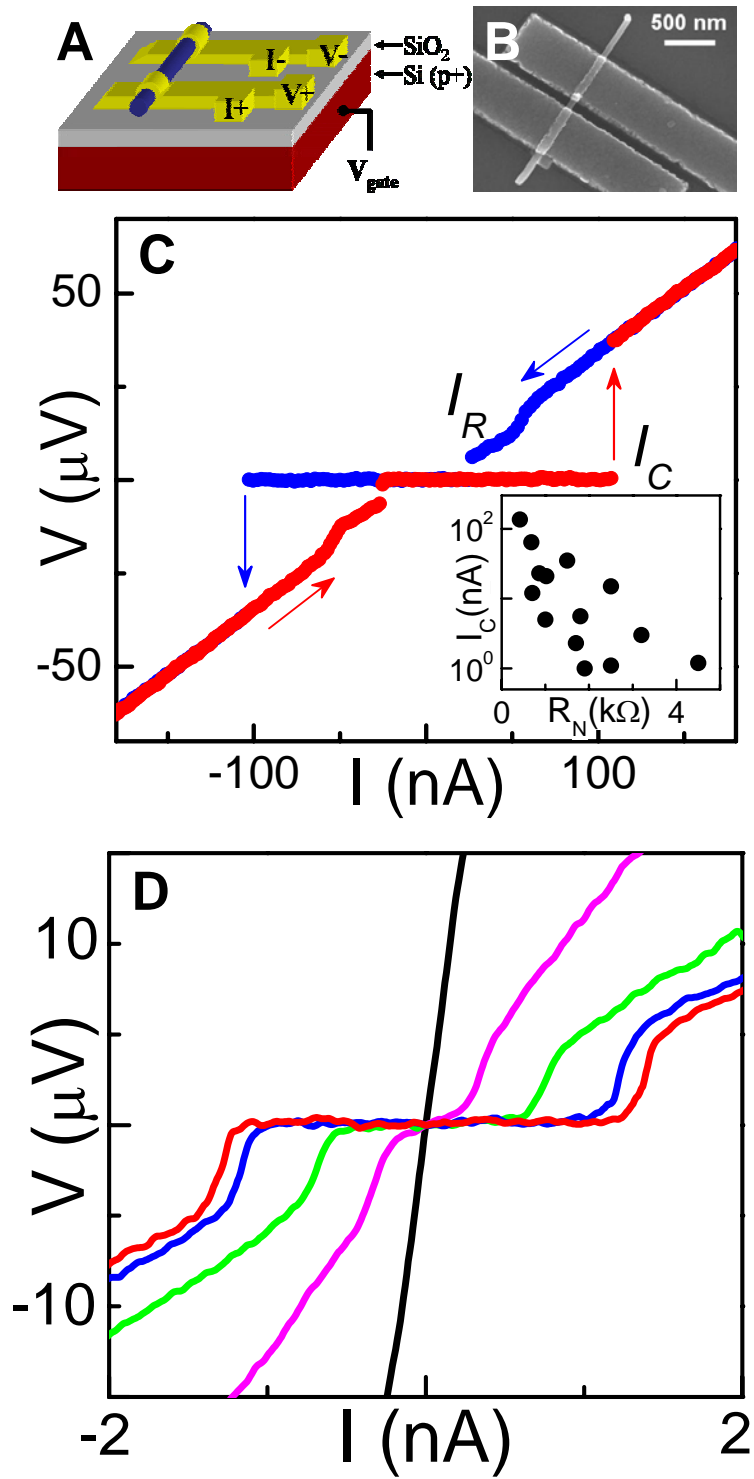

(Y.-J. Doh, *et al.*)
"Tunable Supercurrent Through Semiconductor Nanowires"

# Fig. 2

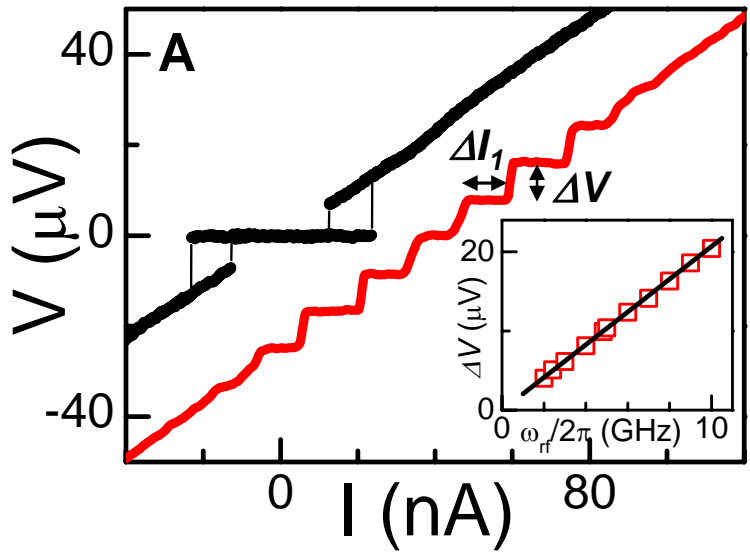

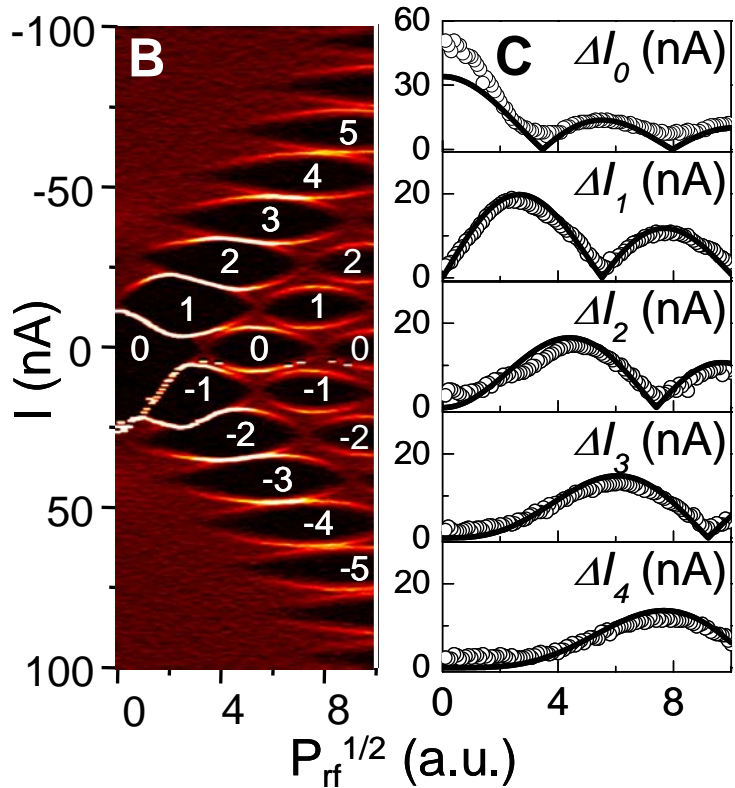

(Y.-J. Doh, *et al.*)
"Tunable Supercurrent Through Semiconductor Nanowires"

# Fig. 3

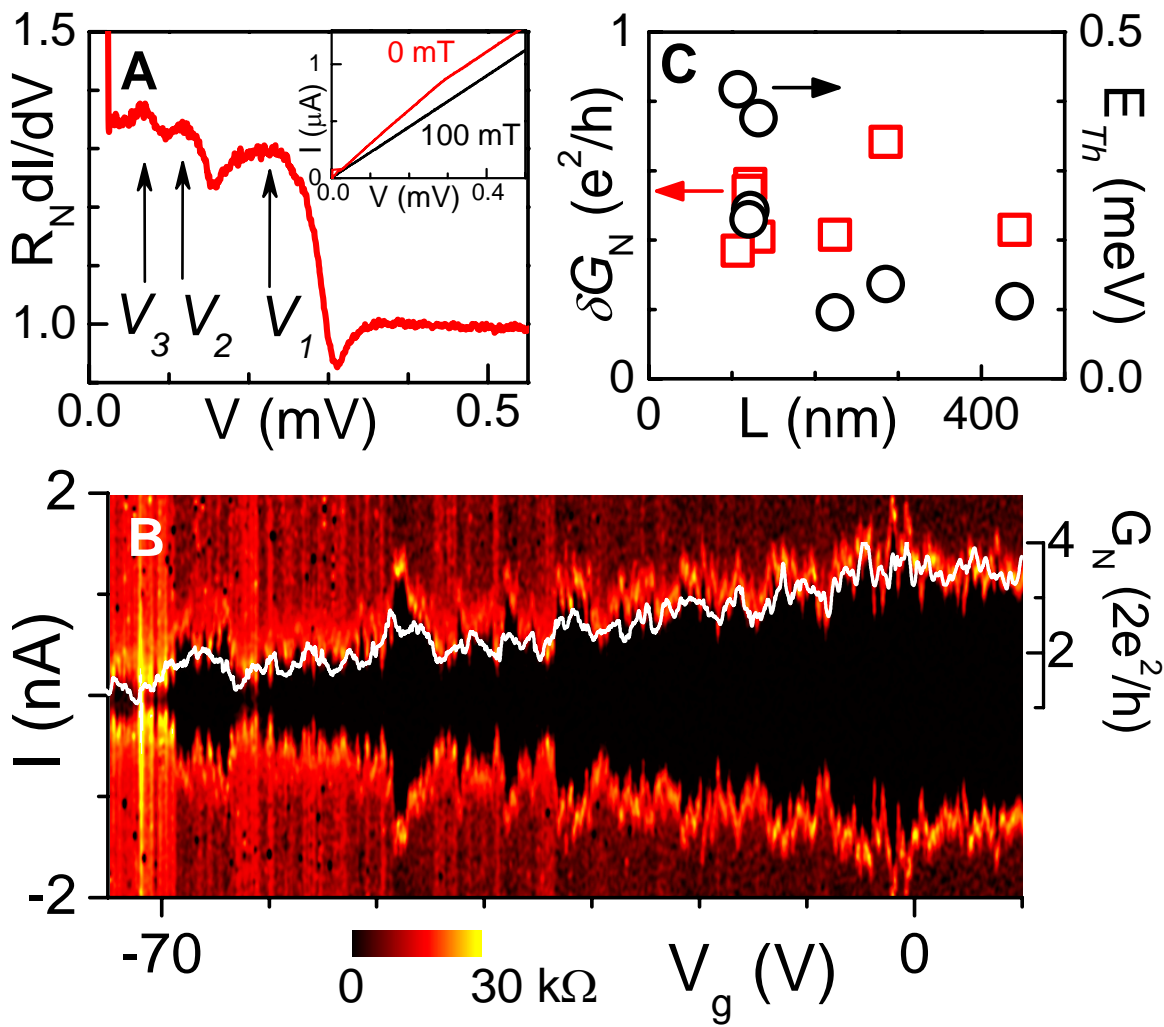

(Y.-J. Doh, *et al.*)

"Tunable Supercurrent Through Semiconductor Nanowires"